\documentclass[12pt,epsf,epsfig]{iopart}
\usepackage{epsfig}
\newcommand \bea{\begin{eqnarray}}
\newcommand \eea{\end{eqnarray}}
\newcommand \beq{\begin{eqnarray}}
\newcommand \eeq{\end{eqnarray}}
\newcommand \ga{\raisebox{-.5ex}{$\stackrel{>}{\sim}$}}
\newcommand \la{\raisebox{-.5ex}{$\stackrel{<}{\sim}$}}

\begin{document}

\title[]{Pairing gaps in Atomic Gases at the BCS-BEC Crossover}

\author{Henning Heiselberg 
\footnote[3]{and Univ. of Southern Denmark (hh@ddre.dk)} 
}

\address{Danish Defense Research Establishment, Ryvangsalle' 1, 
DK-2100 Copenhagen \O , Denmark}
\begin{abstract}
 
Strong evidence for pairing and superfluidity has recently been found
in atomic Fermi gases at the BCS-BEC crossover both in collective
modes and RF excitation energies. It is argued that the scale for the
effective pairing gaps measured in RF experiments is set by the lowest
quasiparticle in-gap excitation energies. These are calculated at the
BCS-BEC crossover from semiclassical solutions to the
Bogoliubov-deGennes equations. 
The strong damping of the radial breathing mode observed in the BCS limit
occur when the lowest quasiparticle excitation energies
coincide with the radial frequency, which indicates that a coupling between
them take place.

\end{abstract}



\section{Introduction}
The recent developments in cold Fermi gases mark another milestone a
century of marvelous discoveries within the field of superfluidity and
superconductivity.  Experiments have established the stability of cold
Fermi gases with strong attractive interactions
\cite{Thomas,Bourdel,Regal,Ketterle,Jochim} and proven the unitarity
limit near Feshbach resonances.  A smooth BCS-BEC crossover is found
in which the Fermi atoms gradually bind into (Bose) molecules near
Feshbach resonances as predicted in Refs. \cite{Eagles,Leggett,Randeria}.  The
collective modes provide strong evidence for superfluidity
\cite{Kinast,Bartenstein}. RF spectroscopy \cite{Chin} clearly show a
peak in the response function that depends on interaction strength,
density and temperature as expected from resonance superfluidity
theory \cite{Levin,Torma}.

 The advantage of the trapped atomic gases is that we can tune the
interactions, densities, temperatures and other trap parameters in a
controlled way and thus explore superfluidity in detail as well as in
general. This insight has already been exploited to describe
some features of pairing in nuclei and neutron stars
\cite{Carlson,HH}.  Furthermore, the pairing gaps $\Delta$ are of
order the Fermi energy $E_F$ around the BCS-BEC crossover and
therefore $\Delta/E_F$ is an order of magnitude larger than in high
temperature superconductors and superfluid $^3$He, and two orders of
magnitude larger than in standard superconductivity (see, e.g,
\cite{Holland}).  Just a few years ago such large pairing gaps was not
considered possible by most of the condensed matter community. In a
uniform system the gap is $\Delta=0.54E_F$ in the unitarity limit
according to Monte Carlo calculations \cite{Carlson}. The observable
transition frequencies are, however, related to the quasiparticle
energies which are considerably smaller than the pairing field in the
center of the trap and therefore in-gap excitations.

 The purpose of this work is to compare the recent experimental results
on RF spectroscopy and collective modes with theoretical calculations.
After a general introduction to pairing and superfluidity in uniform
systems in section II, we address pairing in traps in section III with
special emphasis on the in-gap quasiparticle excitations at the BCS-BEC
crossover, and how they compare to the data of \cite{Chin} in section IV.
In section V we discuss the collective mode frequencies and damping
and the evidence for superfluidity, and finally give a summary.

\section{Pairing in Uniform systems}

The quasiparticle excitations in traps are usually very different
from those in uniform system.  Yet the bulk pairing field is important
for understanding the BCS-BEC crossover, and it is very useful as a
basis for pairing in traps with many particles within the Thomas-Fermi
approximation (TF).

Solving the gap equation at zero temperature for a
Fermi gas interacting through an attractive scattering length $a<0$
gives a pairing gap in the dilute limit, $ak_F\ll -1$, 
\bea
   \Delta = \kappa E_F
   \exp\left[\frac{\pi}{2ak_F}\right] \, .  \label{Gorkov}
\eea
Here, $\kappa=8/e^2$ in standard BCS. Gorkov included
induced interactions which reduces the gap by a factor $\sim 2.2$ 
to $\kappa=(2/e)^{7/3}$
\cite{Gorkov}. In the unitarity limit $k_F|a|\ga 1$, 
which can be reached around the Feshbach resonances, the
gap is of order the Fermi energy \cite{Leggett,HH,Holland}.
Extrapolating (\ref{Gorkov}) to $ak_F\to\pm\infty$ gives a number close to that
found from odd-even staggering binding energies $\Delta=0.49E_F$
calculated by Monte Carlo \cite{Carlson}.

The pairing gap can qualitatively be followed in the crossover model of
Leggett \cite{Leggett} by solving the gap equation
\beq
  1= \frac{2\pi\hbar^2 a}{m} \sum_{\bf k} 
    \left[\frac{1}{\varepsilon_{\bf k}}-\frac{1}{E_{\bf k}} \right] \,.
\eeq
Here, the quasiparticle energy is
$E_{\bf k}=\sqrt{(\varepsilon_{\bf k}-\mu)^2+\Delta^2}$ with
 $\varepsilon_{\bf k}=\hbar^2k^2/2m$.
The chemical potential $\mu$ follows from conservation of 
particle number density
\beq
   n =\sum_{\bf k}\left[1-\frac{\varepsilon_{\bf k}-\mu}{E_{\bf k}} \right] \,.
\eeq
In the dilute (BCS) limit the gap equation leads to the standard BCS gap 
of Eq. (\ref{Gorkov}) - not including the Gorkov correction (see Fig. 1).
The chemical potential is
$\mu=E_F$ and does not include the standard
mean field Hartree-Fock correction $2\pi an/m$ of a dilute gas.

The advantage of the crossover model is that it extends to
the strongly interacting (molecular BEC) limit. Here, the 
pairing gap approaches $\Delta=4E_F/\sqrt{3\pi ak_F}$.
The chemical
potential approaches half of the molecular binding energy $E_b=-\hbar^2/ma^2$
with a BEC mean field corresponding to a molecular scattering length
of $a_m=2a$. Four-body \cite{Petrov}, Monte Carlo calculations 
\cite{Casulleras} and experiments \cite{Ketterle} 
do, however, indicate that $a_m\simeq 0.6a$

On the BCS side ($ak_F<0$) the minimum quasiparticle energy is
$\Delta$ and occur when $k=k_F$. On the BEC side ($ak_F>0$) 
the chemical potential
is negative and the minimum quasiparticle excitation energy is the
quasiparticle energy for ${\bf k}=0$
\beq \label{Leg}
  E = \sqrt{\mu^2+\Delta^2} \,.
\eeq
The chemical potential is observed in the
spin excitation response function \cite{Zoller}.

Monte Carlo calculations \cite{Carlson,Casulleras} of binding
energies, equation of states and pairing gaps agree qualitatively with
the crossover model.  The pairing gap is $\Delta/E_F=0.54$ in the
unitarity limit \cite{Carlson} which is quite close to the
extrapolation of the Gorkov gap
$\Delta^{Gorkov}/E_F=(2/e)^{7/3}\simeq0.49$ but somewhat smaller than
that of the crossover model $\Delta^{Leggett}/E_F=0.69$.

In the BCS limit the critical temperature for the BCS transition is
$T_c=(e^C/\pi)\Delta$, where $C=0.577..$ is Eulers constant.  In
crossover model there are two transition temperatures in the BEC
limit \cite{Randeria}, namely the molecular BEC one at
$T_c=[n/2\xi(3/2)]^{2/3}\pi/m=0.218E_F$ and the molecule dissociation
temperature $T_c^{dissoc}=E_b/\ln(E_b/E_F)^{3/2}$.

\section{Pairing in harmonic oscillator traps}

The various regions of pairing were described in \cite{HM} 
for a dilute system of $N$ atoms with Hamiltonian
\bea  \label{H}
  H &=& \sum_{i=1}^{N}   H_0({\bf r}_i)
  + 4\pi\hbar^2\frac{a}{m} \sum_{i<j} 
     \delta^3({\bf r}_i-{\bf r}_{j}) \,,
\eea 
in a harmonic oscillator (h.o.) potential $H_0({\bf r})={\bf p}^2/2m +
\sum_{k=1}^3 m\omega_k^2 r_k^2/2$.  We shall mention a few relevant
results only before we investigate the strongly interacting limit. 

At least two dimensionless parameters are required to describe this
system even in the spherically symmetric case:
$\omega_1=\omega_2=\omega_3\equiv\omega_0$. These are, e.g., the
number of particles $N$ and the interaction strength $a$ (when
energies are measured in units of $\hbar\omega_0$ and lengths in
$a_{osc}=\sqrt{\hbar/m\omega_0}$).  Several interesting
pairing regions or ``phases'' appear vs. $N$ and $a$.  In contrast,
the binding energies and pairing in an uniform gas at zero
temperature are functions of one parameter only, e.g. $ak_F$.

When the traps contain relatively few atoms, $N\la10^3$, that are weakly
interacting, the mean field does not significantly 
lift the degenerate angular momentum
states $l=n_F,n_F-2,...,1$ or 0, where $n_F=E_F/\hbar\omega=(3N)^{1/3}$,
due to the SU(3) symmetry of the spherical symmetric h.o. potential. 
Consequently, pairing takes place between all these states which
leads to the supergap \cite{HM}
\bea \label{G}
 \Delta= G \equiv\frac{32\sqrt{2n_F}}{15\pi^2}\frac{|a|}{a_{osc}}\hbar\omega_0 
 = \frac{32}{15\pi^2}k_F|a|\hbar\omega_0 \,.
\eea
Here, $k_F=\sqrt{2n_F}/a_{osc}\simeq 1.7N^{1/6}/a_{osc}$ 
is the Fermi wave number in the center of the trap.
For more particles in the trap the stronger mean field cause 
level splitting, which reduce pairing toward single
level pairing, $\Delta_{n_F,l}$, which
displays a distinct shell structure
with h.o. magic numbers. Pairing in nuclei has a similar
shell structure on top of an average gap equal to the supergap
which for constant density scales with the atomic mass number
$A$ as $\Delta\simeq G\simeq5.5{\rm MeV}/A^{1/3}$.
For very large nuclei pairing approaches that in bulk matter
$\Delta \simeq 1.1 {\rm MeV}$.

For stronger interactions pairing also takes place between shells
and the gap increases to  \cite{HM}
\bea \label{GG}
 \Delta= \frac{G}{1-2\ln(e^C n_F)G/\hbar\omega_0} \,.
\eea
This multi-shell pairing region extends up to $2G\ln(e^C n_F)\la \hbar\omega$.

For even stronger interactions the pairing field exceeds
the harmonic oscillator energy in the center of the trap
and the coherence length, $\xi=k_F/\pi m\Delta$, is 
much smaller the TF radius $R_{TF}$ of the cloud.
This spatially inhomogeneous case with a 
strong pairing field $\Delta({\bf r})$ 
can be solved by the Bogoliubov-deGennes equations 
\bea
  E_\eta u_\eta({\bf r}) &=& \quad [H_0+U({\bf r})-\mu]u_\eta({\bf r}) +
  \Delta({\bf r})v_\eta({\bf r})    \,, \nonumber \\
  E_\eta v_\eta({\bf r}) &=& -[H_0+U({\bf r})-\mu]v_\eta({\bf r}) +
  \Delta({\bf r})u_\eta({\bf r})    \,. \nonumber
\eea
Here, $E_\eta$ are the quasiparticle energies, $u_\eta$ and $v_\eta$
the Bogoliubov wave functions, and $U({\bf r})$ the mean field.
The pairing field can be
approximated by the Gorkov gap in the TF approximation \cite{HM}.

Let us first study the spherical symmetric trap.
Tthe Bogoliubov wave function can be
written on the form $u_\eta=u_{nl}(r)Y_{lm}(\theta,\phi)$, where
$(l,m)$ are the angular momentum quantum numbers.
The Bogoliubov-deGennes equations can be solved semiclassically
\cite{Baranov} which leads to a WKB quantization condition
\beq \label{semi}
 (n+\frac{1}{2}) \frac{\pi}{2}= m\int_{R_1}^{R_2} dr
 \frac{\sqrt{E_{n,l}^2-\Delta(r)^2}}{\sqrt{2m(\mu-U(r))-r^2/a^4_{osc}-l(l+1)/r^2}} 
  \,.
\eeq
Here, $n$ is the number of radial nodes 
which in the dilute limit is the h.o. number counted from the Fermi level.
$R_{1,2}$ are the classical turning points. When the pairing
field is strong, i.e. much larger than the quasiparticle excitation
energies, it determines the inner turning point by
$\Delta(R_1)=E_{n,l}$. $R_1$ is then close to the outer turning point
$R_2\simeq R_{TF}$.  The single particle excitations therefore take
place near the surface of the trap where the pairing field is weak.
Furthermore, since $R_2\simeq R_{TF}$ the centrifugal potential can be
neglected when the angular momentum is small, $l\ll n_F$, and we shall
denote this set of energies by $E_n$.

In the dilute BCS limit $\mu\simeq E_F$ and $R_2=R_{TF}=\sqrt{2n_F}a_{osc}$
in Eq. (\ref{semi}), and the quantization condition reduces to
\beq \label{l=0}
 (n+\frac{1}{2}) \frac{\pi}{2}\hbar\omega_0 = \int_{R_1}^{R_{TF}} dr
   \frac{\sqrt{E_n^2-\Delta(r)^2}}{\sqrt{R^2_{TF}-r^2}}  \,,
\eeq
independent of $l$ as long as $l\ll n_F$.
In this expression we can use TF for the pairing field
$\Delta(k_F(r))$ of Eq. (\ref{Gorkov}), i.e. with
$k_F(r)=\sqrt{2n_F(1-r^2/R_{TF}^2)}/a_{osc}$ and $E_F(r)=\hbar^2k_F^2(r)/2m$.
The resulting quasiparticle energies  $E_n$ \cite{HM}
are given by inverting the relation
\begin{equation}\label{largeD}
   k_F|a| \hbar\omega_0 \simeq \frac{E_n}{(n+1/2)}
   \left\{\ln\left[\frac{\kappa(n+1/2)^2\pi^2n_F}{4}
   \frac{(\hbar\omega_0)^3}{E_n^3}\right]\right\}^{-1} \,.
\end{equation}
It is valid above the multishell pairing regime of Eq. (\ref{GG}) and up
to the dense limit $1/(ak_F)\la -1$.

We can furthermore extend the semiclassical model to stronger
interactions and the unitarity limit. As in the Leggett crossover
model we assume that the correlations are included in the pairing
field and that mean field can be included in $\mu$ and in an effective
h.o. potential.  The Bogoliubov-deGennes equations can then be applied
- specifically the semiclassical quantization condition of
Eq. (\ref{semi}).  However, although the crossover model provides a
simple and qualitatively description of the ground state around the
BCS-BCS crossover, it differs from the ground state found by Monte
Carlo calculations and 4-body calculations for both the pairing field
and the chemical potential as mentioned above. We will therefore
include the effect of mean field by replacing the chemical potential
by that calculated by Monte Carlo \cite{Carlson,Casulleras}.  Induced
interactions we include by using the Gorkov result of
Eq. (\ref{Gorkov}) for the pairing field which also is a fair
approximation around the unitarity limit according to Monte-Carlo
calculations \cite{Carlson}.  

Eq. (\ref{semi}) can now be solved in
the Thomas-Fermi approximation.
For example, in the unitarity limit $\Delta(r)=\kappa E_F(r)$
and the attractive mean field is $U=\beta E_F(r)$, where
$\beta\equiv E_{int}/E_{kin}$ is an universal parameter in the
unitarity limit \cite{HH} and is directly related to the chemical potential at 
zero temperature. It has been measured directly from
expansion energies \cite{Thomas,Bourdel,Ketterle,Jochim} and confirms
Monte Carlo calculations $\beta\simeq-0.56$ in the unitarity limit
\cite{Carlson,Casulleras}.  
 The integral in eq.  (\ref{semi}) can be expanded for radii near the surface
and calculated analytically  
\beq\label{hugeD}
   E_n=\left[(n+\frac{1}{2})\frac{\Gamma(7/4)}{\Gamma(5/4)}
    \hbar\omega_0\right]^{2/3} 
    \frac{(\pi\kappa E_F)^{1/3}}{\sqrt{1+\beta}} \,.
\eeq
Note that the form of the mean field in the unitarity limit is such that
it can be included in the chemical potential and h.o. potential and
``renormalizes'' these quantities by factors of $(1+\beta)$ to some power. 
The mean field decreases the TF trap size
$R_{TF}\propto (1+\beta)^{1/4}$, and the chemical potential but increases
the central density. As result the quasiparticle energies increase
slightly.  In Eq. (\ref{hugeD}) the Fermi energy is defined as
$E_F=n_F\hbar\omega_0$ without mean field corrections as also defined
in the experiment of \cite{Chin}.

Also note that the finite size of the system is manifest because
$E_n/E_F$ is not only a function of $ak_F$ but also depends on
$n_F=E_F/\hbar\bar{\omega}$. For example, $E_n/E_F$ scales with
particle number as $\sim N^{-2/9}$ in the unitarity limit.

Around the unitarity limit the semiclassical quantization condition is
solved numerically with the TF Gorkov gap and a mean field correction
$\beta$ averaged over densities in the trap. On the BEC side the
chemical potential $\mu$ should be included in excitations as in
Eq. (\ref{Leg}).  For the lowest quasiparticle excitations near the
surface region, where the density is low, this chemical potential is
within TF simply the molecular binding energy,
$\mu=E_b/2=-\hbar^2/2ma^2$. The quasiparticle energy $E_n$ is
therefore replaced by $\sqrt{E_n^2+(E_b/2)^2}$.  The lowest
quasiparticle energies $E_0$ and $E_1$ including this molecular
binding energy correction are shown in Fig. 1 as function of
$1/(ak_F)$. They reduce to Eqs. (\ref{largeD}) and (\ref{hugeD}) in
the BCS and unitarity limits respectively. On the BEC side the pairing
becomes negligible as compared to the molecular binding energy and
$E\simeq |E_b|/2$.  The quasiparticle energies are much smaller than
the pairing field in the center of the trap and are therefore in-gap
surface excitations.

The Bogoliubov-deGennes equations and semiclassical solutions can
be extended to deformed traps. The quasiparticle energy degeneracy for
$l=0,1,....$ will be split and depend on $\omega_i$.
However, due to energy weighted sumrules we expect that the average
quasiparticle energies can be approximated 
by Eqs. (\ref{largeD}) and  (\ref{hugeD}) when 
$\omega_0$ is replaced by $\bar{\omega}=(\omega_1\omega_2\omega_3)^{1/3}$.
This in concordance with the analysis of the resonance
superfluidity theory which is independent of the symmetry of the cloud
\cite{Torma}.

\vspace{0.cm}
\begin{figure}
\begin{center}
\psfig{file=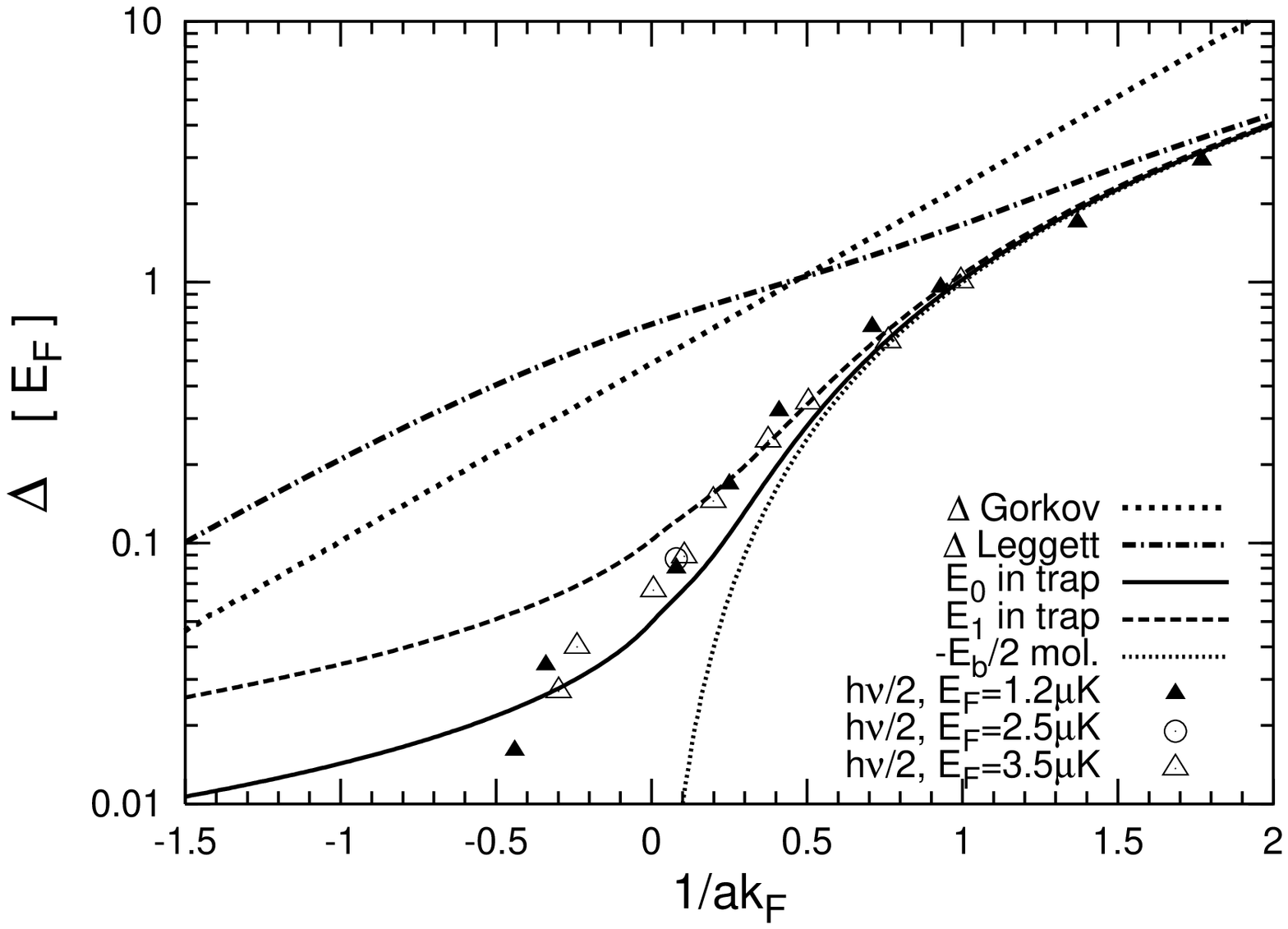,height=14.0cm,angle=-90}
\vspace{.2cm}
\begin{caption}
{Pairing gaps in units of the Fermi energy vs. interaction strength
$ak_F$ in the center of the trap. The full and dashed lines
are the lowest two quasiparticle excitation energies $E_0$ and $E_1$
 (see text).
The ``effective'' pairing gaps $h\nu/2$ measured
in \cite{Chin} at three Fermi energies are shown with symbols.
}
\end{caption}
\end{center}
\label{Grimm}
\end{figure}

\section{Radio Frequency Excitation Spectrum}

For detailed calculations of RF response functions for trapped Fermi
atoms and at finite temperature we refer to recent calculations based on
resonance superfluidity theory \cite{Torma}. Their Bose-Fermion model
divides the system of strongly correlated atoms into molecular bosons
and atoms and introduces effective couplings as well as cutoffs
between these constituents.  They can qualitatively explain the RF
response functions at the various interaction strengths, densities and
temperatures but exceed the effective pairing gaps quantitatively.  We
shall now attempt to understand the RF excitation spectrum in terms of
the quasiparticle energies calculated above at zero temperature.

The ``effective'' pairing gaps reported in \cite{Chin} are defined at
the maximum of the RF response function.  The excitations energies
include the breaking of the pair in states $|1\rangle+|2\rangle$ to an
in-gap quasiparticle in state $|1\rangle$ and an unpaired
quasiparticle in the new spin-state $|3\rangle$. The latter energy is,
however, subtracted on average as it is related to the thermal
peak. The RF excitation spectrum is therefore distributed around the
lowest quasiparticle excitation energies. 
As discussed above we shall  assume that the scale of the RF spectral
weight in a deformed trap 
is given by the lowest quasiparticle
excitations  as given Eqs. (\ref{largeD}) and (\ref{hugeD}) replacing
$\omega_0$ with $\bar{\omega}$.
The distribution is also smeared
by the spectrum of final h.o. states $|3\rangle$ and by finite
temperature effects and damping. The spectral response function is a
sum over quasiparticle excitation energies weighted with a transition
matrix elements. These typically decrease rapidly with transition
energy \cite{Zoller} and therefore the maximum of the response
function is expected to peak at or near the lowest quasiparticle
energies. In the BEC limit the maximum of the response function is
larger than the lowest quasiparticle energy
$|E_b|/2$ by a factor $(4/3)$ only \cite{Chin}.

In Fig. 1 we compare the two lowest quasiparticle excitations $E_0$
and $E_1$ to the effective pairing gaps $h\nu/2$ of Ref. \cite{Chin}
for values of $ak_F$ around the unitarity limit.  $E_n$ are calculated
from Eq. (\ref{semi}), $k_F$ from the central trap densities, and $a$
is calculated assuming that the Feshbach resonances resides at
$B=845G$.  We observe that the effective pairing gaps lie between
lowest quasiparticle energies $E_0$ and $E_1$ as argued above - except
in the dilute BCS limit. The latter departure may be a finite
temperature effect. In the BCS limit the pairing gap and thus $T_c$
decrease toward the temperature of the gas leading to quenching of the gap.
Unfortunately, the temperature can presently not be measured
accurately in traps.

Theoretically the quasiparticle energies and pairing gaps in units of
the Fermi energy are universal function of $ak_F$ around the unitarity
limit when $n_F$ is unchanged.  The effective pairing gaps at three
different Fermi energies obey this scaling to a good approximation.
It should be noted that the exact position of the Feshbach resonance
is important otherwise the scaling is destroyed. By placing the
resonance in $^6$Li around $B\simeq845G$ the effective pairing gaps
scale almost perfectly as seen in Fig. 1.  Best agreement between
experimental and calculated collective modes, that will be discussed
in the following section, is found for a Feshbach resonance residing
around $B\simeq837G$ \cite{Hu}. For an even lower Feshbach resonance
systematic deviations appear between the effective pairing gaps and
$E_F$.

We emphasize that the pairing gaps are calculated from the
semiclassical quantization condition of Eq. (\ref{semi}) which is
based on the Bogoliubov-deGennes equations derived from the gap
equation in the crossover mode. 
However, although 
the crossover model provides a simple and qualitatively description
of the ground state around the BCS-BCS crossover, it differs from the ground
state found by Monte Carlo calculations and 4-body calculations for both
the pairing field and the chemical potential as mentioned
above.
The effect of mean fields were included by replacing 
the chemical potential by that calculated by Monte Carlo. Induced interactions
were included by using the Gorkov result of Eq. (\ref{Gorkov}) for the
pairing field which also
is a fair approximation around the unitarity limit
according to Monte-Carlo calculations \cite{Carlson}.

\section{Collective Modes}

The hydrodynamic and superfluid collective mode frequencies are
unfortunately identical and we have to rely on other observables to be
able to distinguish them and prove, e.g. superfluidity. We shall first
discuss the collective modes for a very large number of particles in a
trap and/or with strong interactions. Subsequently, we discuss the
effect of a finite number of particles in a trap and damping with and
without superfluidity.

The hydrodynamic and superfluid collective modes can be calculated in
general for a polytropic equation of state: $P\propto
n^{1+\gamma}$. Here, the polytropic power is $\gamma=1$ in a dilute
interaction dominated BEC, whereas an ideal Bose gas in the normal
state has $\gamma=2/3$ under adiabatic conditions. A dilute gas of
Fermi atoms also has $\gamma=2/3$ in both the hydrodynamic and
superfluid limits.  Both a Fermi gas \cite{HH} and a BEC \cite{Cowell}
has $\gamma=2/3$ in the strongly interacting (unitarity) limit. The
effective power $\gamma$ has been calculated at the BCS-BEC crossover
for the Leggett model \cite{mode,Hu} and by Monte Carlo \cite{Carlson}
and varies between $\gamma\sim 0.5-1.3$.
 
A spherical symmetric h.o. trap with a polytropic equation of state
has collective mode frequencies \cite{mode}
$\omega^2=\omega_0^2(l+2n[\gamma(n+l+1/2)+1])$, where $l$ is the
angular momentum and $n$ the number of radial nodes.  In comparison
the collective modes in the collisionless limit are those of a free
particle: $\omega/\omega_0=2n+l$, when its mean free path exceeds the
size of the cloud.

In an axial symmetric trap: $\omega_1=\omega_2\equiv\omega_\perp$ and
$\omega_3=\lambda\omega_\perp$, the resulting breathing modes are the
coupled monopole and quadrupole $m=0$ modes \cite{Cozzini}.  For the
very cigar shaped traps $\lambda\ll 1$ used in \cite{Kinast} and
\cite{Bartenstein} the coupled modes become the radial
\beq \label{rad}
   \omega_{rad}=\sqrt{2(\gamma+1)} \,\omega_\perp \,,
\eeq 
and axial
\beq \label{ax}
   \omega_{ax}=\sqrt{3-(\gamma+1)^{-1}}\omega_3 \,,
\eeq
modes.
Taking the effective power $\gamma$ from the Leggett model at the
BCS-BEC crossover good agreement is found \cite{mode,Hu} with the
experiments of \cite{Kinast} and \cite{Bartenstein} for the axial and
and for the radial modes of  \cite{Kinast}.
The radial mode in \cite{Bartenstein} differs and a transition or ``break'' 
is observed 
around the break point $B\simeq 910G$ which is accompanied by strong damping.

In the unitarity limit $x=1/(ak_F)=0$ scaling predicts that
$\gamma=2/3$ as is also found in 
the axial and radial mode of \cite{Kinast} and in the axial mode of
\cite{Bartenstein}. Furthermore, it follows from
Eqs. (\ref{rad}) and (\ref{ax}) that as function of $x=1/(ak_F)$ their
slopes are: $\omega_{rad}'/\omega_\perp=\gamma'\sqrt{3/10}$ and
$\omega_{ax}'/\omega_3=\gamma'(3/5)^{3/2}/4$ in the unitarity limit.  
The slope of $\gamma$
and $\beta$ can at $x=0$ be related as \cite{mode}
\bea
  \gamma' = \frac{\beta'}{6(1+\beta)} \,.
\eea 
The Monte Carlo calculations \cite{Carlson,Casulleras},
the crossover model and the LOCV model
\cite{Carlson,HH} all find $\beta'(x=0)\simeq
-0.10$ and $\beta(x=0)\simeq -0.56$ (the crossover model does, however, give
$\beta(x=0)=-0.42$ which probably is due to the lack of
Hartree-Fock energy corrections).
Thus we obtain
$\gamma'(x=0)\simeq 0.40$. The resulting slopes of the collective frequencies
$\omega_{rad}'$ and $\omega_{ax}'$ 
are also compatible with experiment \cite{Kinast,Bartenstein} 
in the unitarity limit.

The measured damping of the modes is not compatible with hydrodynamics.  As
pointed out in \cite{Kinast} the damping rate increase with increasing
temperature whereas the opposite is expected in hydrodynamics. In a
superfluid the condensate is gradually depleted as the temperature
increases and coupling between the normal and superfluid components
increase damping as observed. The damping in \cite{Bartenstein}
peaks at the transition which together with the abrupt transition in
frequency indicates a superfluid to collisionless transition rather
than a smooth hydrodynamic to collisionless transition.

The damping rate from collisional damping in a normal gas
can be estimated from the semi-quantitative form for the
transition between collisionless and hydrodynamic limit \cite{PS,Gensemer}
\bea
 \omega^2 =  \omega^2_C - \frac{ \omega^2_C-\omega^2_H}{1-i\omega\tau_{coll}}
   \,.
\eea
For the radial mode the hydrodynamic frequency is 
$\omega_H\simeq\sqrt{10/3}\omega_\perp$ whereas the collisionless is 
$\omega_C\simeq 2\omega_\perp$.
The relaxation processes are 
expressed in terms of a collision time $\tau_{coll}$.
The maximal damping 
of the collective mode $Im(\omega)/\omega$ occurs between the
hydrodynamic and collisionless limits for $\omega\tau_{coll}=1$
and is: $Im(\omega)\simeq 0.09\omega_\perp$, for the radial mode. 
The damping rates in
\cite{Bartenstein} and \cite{Thomasprelim} are considerably larger around the
break points, and can thus not be caused by collisional damping alone.  

Alternatively, the enhanced damping and break point could be
caused by a superfluid to collisionless transition. However, in the center
of the trap the critical temperature $T_c=0.28E_F\exp(\pi x/2)$ is 
at the break point $x\simeq -0.5$ much larger
than the gas temperatures, $T\la 0.03E_F$. 
Thus only the low density surface layer is not in the superfluid phase.
Estimates of the effective scattering cross sections
indicate that the gas is collisionless in the normal phase at the very low
temperatures present in the experiments of 
\cite{Bartenstein} and \cite{Kinast,Thomasprelim}.

It was pointed out in Ref. \cite{Bartenstein} that the lowest
quasiparticle excitation energy is comparable to the collective
energy $\hbar\omega_\perp$ of the radial mode 
at their transition.  An
interesting coupling between the single particle states and the
collective mode may therefore take place which is special for a finite
system. This coupling can be studied in the RPA equations which
describe how the collective modes are build up of quasiparticle
states.  The collective modes are calculated for dilute and spherical
h.o. traps in \cite{Bruun} by solving the RPA equations using the
quasiparticle states and energies as input. When interactions are
weak such that $2G\ln(e^C n_F)\la\hbar\omega_0$, the collective modes
are dominated by the pairing gap and the h.o. shell structure. For
stronger interactions the collective modes can be calculated from RPA
and they approach those of a superfluid.

 Though the collective modes for both the dilute system and the unitarity
regime typically are of order the harmonic oscillator frequency, the underlying
quasiparticle spectrum is very different. In the dilute limit the
pairing gap and the lowest excitation energies can be far below
collective energies. This is similar to the situation in atomic nuclei
where pairing energies are of order 1MeV whereas giant resonances lie
around 10-20MeV. In contrast, for trapped atoms in the unitarity
regime the quasiparticle excitation energies $E_n$ lie well above
$\hbar\bar{\omega}$. Towards the BCS limit, however, $E_0$ approaches
$\hbar\omega_{rad}$. It follows from the RPA equations
that the collective frequency both decreases and is damped with respect
to a system with an infinite number of particles. This is in
qualitative agreement with the experiment of Ref. \cite{Bartenstein}
on the BCS side. 

 The observed transition or break in the radial frequency occurs at $B=910G$
\cite{Bartenstein} for both $\nu_\perp=750$Hz and 2.4kHz, i.e. for the same
scattering length but for two different $k_F$ and radial frequencies. However,
we find that in both cases the resulting radial frequency and lowest quasiparticle energy almost coincide at the break point, i.e.
\bea \label{break}
  \hbar \omega_{rad} \simeq 2 E_0  \,.
\eea
Note that  $E_F$ and $k_F$ do not scale linearly with $\nu_\perp$ 
as does $\omega_{rad}$. It is thus a coincidence that 
Eq. (\ref{break}) is fulfilled for
the same scattering length and therefore at the same B for the two different
$\nu_\perp$. 
In the experiment of Ref. \cite{Kinast} both the trap frequencies are larger
than in \cite{Bartenstein} and one can calculate that the break should occur
around $B\simeq 980G$ according to the condition in Eq. (\ref{break}).
Preliminary data \cite{Thomasprelim} finds that damping does increase at 
such magnetic fields but that a possible break point is situated 
at even higher magnetic fields. 

More precise measurements of $T_c$ and $T$ are required before the precise
phase, damping mechanisms and break points can be determined.

\section{Summary and Outlook}

We have argued that the recent experiments on RF spectroscopy
\cite{Chin} and collective modes \cite{Kinast,Bartenstein} give strong
evidence for superfluidity in traps with Fermi atoms. The results
compare very well with theoretical calculations \cite{Torma,mode,Hu}
and the quasiparticle excitation energies discussed above as function
of interaction strength, density and temperature. 

There are, however, some details and deviations between theory and
experiments that needs further investigation. The effective pairing
gaps of Ref. \cite{Chin} only partially obey the scaling with $E_F$.
The radial mode collective frequency of Ref. \cite{Bartenstein}
violate the scaling result in the unitarity limit and undergo a
transition.  Better control of the temperature and the exact position
of the Feshbach resonance as well as experiments at several densities
and temperatures are all desirable.  These are necessary for a
detailed check of the scaling with $ak_F$ around the unitarity limit
and to pin down transitions and critical temperatures between
superfluid and normal phases, collisionless and hydrodynamic dynamics.

In this respect we can appreciate the great advantage of atomic gases,
namely the great number of tunable parameters such as the number of
particles, densities, interaction strengths, temperatures, trap
deformation, number of spin states, etc.  They therefore hold great
promise for a more general understanding of pairing phenomena atomic
Fermi gases but also in solids, metallic clusters, grains, nuclei,
neutron stars, quark matter, etc.

{\bf Acknowledgements}: Discussions with J. Thomas
and R. Grimm are gratefully acknowledged.


\end{document}